\documentclass[conference]{IEEEtran}

\usepackage{cite}
\usepackage{amsmath,amssymb,amsfonts}
\usepackage{array}
\usepackage[noend]{algpseudocode}
\usepackage[noend,ruled,vlined,linesnumbered]{algorithm2e}
\usepackage{graphicx}
\usepackage{textcomp}
\usepackage{xcolor}
\usepackage{flushend}
\usepackage{multirow}
\usepackage{booktabs}
\usepackage{tabularx}
\usepackage{ragged2e}
\usepackage[final]{microtype}
\usepackage{comment}
\usepackage[alpha]{mdpn} 
\ifCLASSOPTIONcompsoc
  \usepackage[caption=false,font=normalsize,labelfont=sf,textfont=sf]{subfig}
\else
  \usepackage[caption=false,font=footnotesize]{subfig}
\fi

\newcolumntype{L}{>{\RaggedRight\arraybackslash}X} 

\def\BibTeX{{\rm B\kern-.05em{\sc i\kern-.025em b}\kern-.08em
    T\kern-.1667em\lower.7ex\hbox{E}\kern-.125emX}}

\graphicspath{{./images/}}
\begin{document}

\title{DRL-based Slice Placement Under Non-Stationary Conditions}

\author{\IEEEauthorblockN{Jose Jurandir Alves Esteves\IEEEauthorrefmark{1}\IEEEauthorrefmark{2}, Amina Boubendir\IEEEauthorrefmark{1}, Fabrice Guillemin\IEEEauthorrefmark{1} and Pierre Sens\IEEEauthorrefmark{2}}
\IEEEauthorblockA{\IEEEauthorrefmark{1}Orange Labs, France} \IEEEauthorrefmark{2}Sorbonne Universit\'e / CNRS / Inria, LIP6, France\\ \{josejurandir.alvesesteves, amina.boubendir, fabrice.guillemin\}@orange.com, pierre.sens@lip6.fr }

\maketitle

\begin{abstract}
We consider online learning for optimal network slice placement under the assumption that slice requests arrive according to a non-stationary Poisson process. We propose a framework based on Deep Reinforcement Learning (DRL) combined with a heuristic to design algorithms. We specifically design two pure-DRL algorithms and two families of hybrid DRL-heuristic algorithms. To validate their performance, we perform extensive simulations in the context of a large-scale operator infrastructure. The evaluation results show that the proposed hybrid DRL-heuristic algorithms require three orders of magnitude of learning episodes less than pure-DRL to achieve convergence. This result indicates that the proposed hybrid DRL-heuristic approach is more reliable than pure-DRL in a real non-stationary network scenario. 
\end{abstract}


\section{Introduction}

The promise of {network} Slicing is to enable a high level of customization of network services in future networks (5G and beyond) leveraged by  virtualization and software defined networking techniques. These key enablers  transform telecommunications networks into programmable platforms capable of offering virtual networks enriched by Virtual Network Functions (VNFs) and IT resources tailored to the specific needs of certain customers (e.g., companies) or vertical markets (automotive, e-health, etc.)\cite{3GPP,etsi}.

From an optimization theory perspective, the Network Slice Placement problem can be viewed as a specific case  of Virtual Network Embedding (VNE) or VNF Forwarding Graph Embedding (VNF-FGE) problems \cite{survey_vnf_ra_2016}. It is then generally possible to formulate Integer Linear Programming (ILP) problems \cite{netsoft_2020}, which, however, turn out to be $\mathcal{NP}$-hard \cite{vne_np_hardness} with very long convergence time. 

With regard to network management, there are specific characteristics related to network slicing: slices are expected to share resources and coexist in a large and distributed infrastructure. Moreover, slices have a wide range of requirements in terms of resources, quality objectives and lifetime. In practice, these characteristics bring additional complexity as the placement algorithms need to be highly scalable with low response time even under varying network conditions. 

As an alternative to optimization techniques and the development of heuristic methods, Deep Reinforcement Learning (DRL) has recently been used in the context of VNE and Network Slice Placement \cite{p1,p2, p5, p3, p4, p8, sch_cnsm_2020,hab_cnsm_2020}. 

DRL techniques are seen as very promising since they allow, at least theoretically, to learn optimal decision policies only based on experience \cite{sutton2018reinforcement}. However, from a practical point of view, especially in the context of non-stationary environments, ensuring that a DRL agent converges to an optimal policy is still a challenge. 

As a matter of fact, when the environment is continually  changing the rules, the algorithm has trouble in using the  acquired knowledge to find optimal solutions. The usage of the DRL algorithm in a online fashion can then become impractical. In fact, most of the existing works applying DRL to solve the Network Slice Placement or VNE problem assume a stationary environment, i.e., with static network load. However, traffic conditions in networks are basically non-stationary with daily and weekly variations and subject to drastic changes (e.g., traffic storm due to an unpredictable event).   

To cope with traffic changes, we propose in the present paper to extend a hybrid DRL-heuristic algorithm we have recently introduced in \cite{HA_DRL_TNSM} (namely Heuristically Assisted DRL, HA-DRL) to evaluate the performance under non-stationary network loads. We apply this strategy to a fully online learning scenario with time-varying network loads to show how this strategy can be used to accelerate and stabilize the convergence of DRL techniques when applied to the Network Slice Placement problem.

The  contributions of the present paper are threefold: 
\begin{enumerate}
    \item We propose a network load model to network slice infrastructure conditions with time-varying network loads;
    \item We propose a framework combining Advantage Actor Critic and a Graph Convolutional Network (GCN) for conceiving DRL-based algorithms adapted to the non-stationary case;
    \item We show how the DRL learning process can be accelerated by using the proposed HA-DRL technique to control the algorithms convergence.
\end{enumerate}

The organization of this paper is as follows: In Section~\ref{sec:sota}, we review related work. In Section~\ref{sec:network_model}, we describe the Network Slice Placement problem modeling. The learning framework for slice placement optimization is described in Section~\ref{sec:drl_proposal}. The adaptation of the pure-DRL approaches and its control by using heuristic is introduced in Section~\ref{sec:aidedDRL}. 

The experiments and evaluation results are presented in Section~\ref{sec:evaluation}, while conclusions and perspectives are presented in Section~\ref{sec:conclusion}.

\section{Related Work Analysis \label{sec:sota}}

We review, in this section, recent studies on DRL-based approaches for  network slice placement. The reader interested in a more detailed and comprehensive discussion of those works cited in the present paper may refer to \cite{HA_DRL_TNSM}.
\subsection{On Pure-DRL approaches for slice placement\label{sec:ml_based}}

There are only a few recent works on DRL for network slice placement and VNE related problems in the literature and the majority of them are pure-DRL approaches \cite{p1,p2, p5, p3, p4, p8, sch_cnsm_2020,hab_cnsm_2020}. In those works, only the knowledge acquired by the learning agent via training is used as a basis for taking placement decisions. The drawback of this approach is that the learning agent needs extensive exploration of the state and action spaces to learn an appropriate policy for decision making; such a process  takes a lot of time during which the agent  takes bad placement decisions; this leads to rejection of slices and bad performance.

 Furthermore, existing works consider static or fixed average network load regimes in which network slices arrive and exit the system always at  fixed rates. In reality, however, network load conditions  vary over time as the demand for network services depend on many  factors (unpredictable events, day-night variations, etc.). In this kind of non-stationary environment, the learning agent might have trouble in using its knowledge acquired via training as the rules of the environment are constantly changing. 
 
\subsection{On Hybrid DRL-heuristic approaches for slice placement \label{sec:hybrid}}

Two recent works on network slice placement and VNE have been proposed  and combine DRL with heuristic methods to speed up the convergence and to increase the reliability of DRL algorithms \cite{quang2019deep,rkhami2021learn}. However, these hybrid DRL-heuristic approaches have some drawbacks explained in more details in \cite{HA_DRL_TNSM}. In particular, the approach proposed in \cite{quang2019deep} adopts an infinite action space formulation that adds some overhead reducing the applicability of the algorithm, and the approach proposed in \cite{rkhami2021learn} strongly depends on the quality of the initial solution provided by a heuristic. Moreover, these two works do not consider the non-stationarity assumption discussed in the present paper. 

In \cite{HA_DRL_TNSM}, we proposed to adapt heuristically accelerated reinforcement learning  to the network slice placement problem. The approach is inspired by \cite{bianchi2008accelerating} and  addresses the shortcomings of \cite{quang2019deep} and \cite{rkhami2021learn}. In this paper, we extend this approach by including the non-stationarity assumption. 

\subsection{On AI/ML approaches considering dynamic network load\label{sec:dynamic_nw_loads}}

A recent body of research considers dynamic network load scenarios when applying AI/ML-based approaches to support optimization of network slice life cycle management. For instance, the authors of \cite{deepcog} propose a deep learning-based data analytics tool to predict dynamic network slice traffic demands to help avoid SLA violations and network over-provisioning. The paper \cite{load_forecasting} also adopts neural networks to predict network slice traffic demands but, in this case, to perform proactive resource provisioning and congestion control.

To the best of our knowledge, the only paper considering placement optimization using DRL in dynamic network load scenarios is \cite{new_1}.

The authors propose a Double Deep Q Network (DDQN) algorithm for re-optimizing an initial VNF placement. They consider that the network load changes periodically with a time cycle $T$. They then separate time cycle T in time intervals $\Delta t$, and train a DDQN model to specifically take charge of VNF placement re-optimization in each time interval $\Delta t$. Despite its originality, the approach proposed in \cite{new_1} presents two drawbacks: 1) it depends on offline learning, which is  not applicable to online optimization scenarios; 2) it does not use DRL to optimize placement directly as the DRL algorithm selects the region of the network to re-optimize and delegates the optimization to threshold policy procedure. The heuristic calculation of placement decisions can lead to sub-optimal solutions. Contrary to  \cite{new_1}, the present contribution is applicable to offline and online learning and  directly learns a placement optimization policy.  

\section{Network Slice Placement Optimization Problem \label{sec:network_model}}

This section presents the various elements composing the model for slice placement. Slices are placed on a substrate network, referred to as Physical Network Substrate (PSN) and described in Section \ref{sec::psn_model}. Slices give rise to as  Network Slice Placement Requests (Section \ref{sec:nspr_model}), generating a  network load defined in Section \ref{sec:network_load_modeling}. The optimization problem is formulated in Section \ref{sec:nsp_problem_statement}.

\subsection{Physical Substrate Network Modeling \label{sec::psn_model}}

The Physical Substrate Network (PSN) is composed of  the infrastructure resources, namely  IT resources (CPU, RAM, disk, etc.) needed for supporting  the Virtual Network Functions (VNFs) of network slices  together  with the transport network, in particular Virtual Links (VLs) for interconnecting the VNFs of slices. 

The PSN  is divided into three components: the Virtualized Infrastructure (VI) corresponding to IT resources, the Access Network (AN), and the Transport Network (TN). 

The Virtual Infrastructure (VI) hosting IT resources is  the set of Data Centers (DCs) interconnected by network elements (switches and routers). We assume that data centers are distributed in Points of Presence (PoP) or centralized (e.g., in a big cloud platform). As in  \cite{slim2018close}, we define three types of DCs with different capacities: Edge Data Centers (EDCs) close to end users but  with small resources capacities, Core Data Centers (CDCs) as regional DCs with medium resource capacities, and Central Cloud Platforms (CCPs) as national DCs with big resource capacities.

We consider that slices are rooted so as to take into account the location of those users of a slice. We thus introduce an Access Network (AN) representing User Access Points (UAPs) such as Wi-Fi APs, antennas of cellular networks, etc. and Access Links. Users access  slices  via one UAP, which may change during the life time of a communication by a  user (e.g., because of mobility). 

The Transport Network (TN) is the set of routers and transmission links needed to interconnect the different DCs and the UAPs. The complete PSN is modeled as a weighted undirected graph $G_s = (N, L)$ with parameters described in Table \ref{tab::physical_substrate_network}, where $N$ is the set of physical nodes in the PSN, and $L \subset \{(a, b) \in N \times N \wedge a\neq b\}$ refers to a set of substrate links. Each node has a type in the set $\{$UAP, router, switch, server$\}$. The available CPU and RAM capacities on each node are defined  as $cap^{cpu}_n \in \mathbb{R}$, $cap^{ram}_n \in \mathbb{R}$ for all $n \in N$, respectively. The available bandwidth on the links are defined as $cap^{bw}_{(a,b)} \in \mathbb{R}, \forall (a,b) \in L$.
      
\begin{table}[hbtp]
\caption{PSN parameters \label{tab::physical_substrate_network}}
\begin{tabular}{@{}cc@{}}
\toprule                 
\textit{\textbf{Parameter}}                          & \textit{\textbf{Description}}             \\ \midrule
$G_s = (N,L)$       & PSN graph \\
$N$               & Network nodes  \\
$S \subset N$     & Set of servers \\
$DC$             & Set of data centers                       \\
$S_{dc} \subset S$, $\forall dc \in DC$ & Set of servers in data center $dc$        \\
 $SW_{dc}, \ \forall dc \in DC$   & Switch of of data center $dc$  \\
$L = \{(a,b) \in N \times N \wedge a \neq b\}$ & Set of physical links                     \\
$cap^{bw}_{(a,b)} \in \mathbb{R}, \forall (a,b) \in L$ & Bandwidth capacity of  link $(a,b)$ \\
$cap^{cpu}_s \in \mathbb{R}, \forall s \in S$        & available CPU capacity on server $s$                \\
$M^{cpu}_s \in \mathbb{R}, \forall s \in S$        & maximum CPU capacity of server $s$                \\
$cap^{ram}_s \in \mathbb{R}, \forall s \in S$        & available RAM capacity on server $s$ \\
$M^{ram}_s \in \mathbb{R}, \forall s \in S$        & maximum RAM capacity of server $s$ \\
$M^{bw}_s \in \mathbb{R}, \forall s \in S$        & maximum outgoing bandwidth from $s$ \\ \bottomrule
\end{tabular}
\end{table}

\subsection{Network Slice Placement Requests Modeling \label{sec:nspr_model}}

We consider that a slice is a chain of  VNFs to be placed and connected over the PSN. VNFs of a slice are grouped into a request, namely a Network Slice Placement Request (NSPRs), which has to be placed on the PSN.  A  NSPR is  represented as a weighted undirected graph  $G_v = (V, E)$, with parameters described in Table~\ref{tab::nspr_parameters}, where $V$ is the set of VNFs in the NSPR, and $ E \subset \{(\bar{a}, \bar{b}) \in V \times V \wedge \bar{a} \neq \bar{b}\}$ is a set of VLs to interconnect the VNFs of the slice . The CPU and RAM requirements of each VNF of a NSPR are defined as $req^{cpu}_{v} \in \mathbb{R}$ and $req^{ram}_{v} \in \mathbb{R}$ for all $v \in V$, respectively. The bandwidth required by each VL in a NSPR is given by $req_{(\bar{a},\bar{b})}^{bw} \in \mathbb{R}$  for all  $(\bar{a},\bar{b}) \in E$.

\begin{table}[hbtp]
\centering
\caption{NSPR parameters \label{tab::nspr_parameters}}
\begin{tabular}{@{}cc@{}}
\toprule
\textit{\textbf{Parameter}}                                          & \textit{\textbf{Description}}                   \\ \midrule
$G_v = (V,E)$                                                                  & NSPR graph                         \\
$V$                                                                  & Set of VNFs of the NSPR                         \\
$E=\{(\bar{a},\bar{b}) \in N \times N \wedge \bar{a} \neq \bar{b}\}$ & Set of VLs of the NSPR                          \\
$req^{cpu}_{v} \in \mathbb{R}$                                         & CPU requirement of VNF $v$                      \\
$req^{ram}_{v} \in \mathbb{R}$                                         & RAM requirement of VNF $v$                      \\
$req_{(\bar{a},\bar{b})}^{bw} \in \mathbb{R}$                          & Bandwidth requirement of VL $ (\bar{a},\bar{b})$\\ \bottomrule
\end{tabular}
\end{table}

\subsection{Network Load Modeling \label{sec:network_load_modeling}}

We consider the case when the load offered by NSPRs is time varying. We specifically assume that there are several classes of NSPRs and two sets of classes. A first set is composed of NSPR classes (referred to as static) with constant arrival rates, creating a background traffic. A second set is composed of  NSPR classes, with time-varying arrival rates so as to reflect some volatility in the NSPRs. Those NSPRs are said dynamic.

\subsubsection{Network Load for Static NSPR Classes}
Let $J$ be the set of resources in the network (i.e., CPU, RAM, bandwidth). Let $\mathcal{K}_{c} \subset \mathbb{N}$ be the set of static NSPR classes. We compute the load generated by arrivals of NSPRs of class $k$ in $K_{c}$ for resource $j$ in $J$ as in \cite{farah2}:
\begin{equation}
    \rho^{k}_{j} = \frac{1}{C_j}\frac{\lambda^{k}}{\mu^{k}}A^{k}_{j} ,
    \label{eq:static}
\end{equation}
where $C_j$ is the total capacity of resource $j$, $A^k_j$ is the number of resource units requested by an NSPR of class $k$, $\lambda^{k}$ is the average arrival rate for an NSPR of class $k$ and $1/\mu^{k}$ is the average lifetime of an NSPR of class $k$.

\subsubsection{Network Load for Dynamic NSPR Classes}

Let $\mathcal{K}_{\Delta} \subset \mathbb{N}$ be the set of dynamic NSPR classes. We consider a periodic average arrival rate $\lambda^{k}(t)$ for class $k$ in $\mathcal{K}_{\Delta}$ given by 
\begin{equation}
    \lambda^{k}(t) = f^{k} \sin^{2}\left(\frac{\pi t}{ \tau^{k}}\right)
    \label{eq:dynamic}
\end{equation}
where $\tau^{k}$ is the period of $\lambda^{k}(t)$ in time units and $f^{k}$ is a parameter used to control the amplitude of $\lambda^{k}(t)$. We then adapt Eq.~\eqref{eq:static} to compute the network load for dynamic NSPR class $k$ and resource $j$ as $\rho^{k}_{j}(t) = \frac{1}{C_j}\frac{\lambda^{k}(t)}{\mu^{k}}A^{k}_{j}$. It is worth noting that to preserve $\rho^{k}_{j}(t)$ in the $[0,1]$ interval, $f^{k}$ must be between $0$ and $\frac{C_j \mu^{k}}{A^{k}_{j}}$.

\subsubsection{Global Network Load}
Finally, we define the global network load $\rho_{j}(t)$ for each resource $j$ in $J$ in the simulated time instant $t$ as the sum of the network loads generated by static NSPR classes ($K_{c}$) and dynamic NSPR classes ($K_{\Delta}$), that is, 
\begin{equation}
    \rho_{j}(t) = \sum_{k \in K_{\Delta}}\rho^{k}_{j}(t) + \sum_{k \in K_{c}}\rho^{k}_{j}
    \label{eq:global}
\end{equation}
It is worth noting $ 0 \leq \rho_j(t) \leq 1$. 

\subsection{Network Slice Placement Optimization Problem Statement \label{sec:nsp_problem_statement}}


\begin{itemize}
    \item \textit{Given:} a NSPR graph $G_v = (V, E)$ and a PSN graph $G_s = (N, L)$,
    \item \textit{Find:} a mapping $G_v \to  \bar{G}_s =(\bar{N},\bar{L})$, $\bar{N} \subset N$, $\bar{L} \subset L$,
    \item\textit{Subject to:} the VNF CPU requirements $req^{cpu}_v, \forall v \in V$, the VNF RAM requirements $req^{ram}_v, \forall v \in V$, the VLs bandwidth requirements $req^{bw}_{(\bar{a},\bar{b})}, \forall (\bar{a},\bar{b}) \in E$, the server CPU available capacity $cap^{cpu}_s, \forall s \in S$, the server RAM available capacity $cap^{ram}_s, \forall s \in S$, the physical link bandwidth available capacity $cap^{bw}_{(a,b)}, \forall (a,b) \in L$.
    \item \textit{Objective: } maximize the network slice placement request acceptance ratio, minimize the total resource consumption and maximize load balancing.
\end{itemize}

\section{Learning framework for Network Slice Placement Optimization  \label{sec:drl_proposal}}

We describe in this section the machine learning framework used to solve the optimization formulated in Section~\ref{sec:network_model}. We adopt the same approach as in \cite{HA_DRL_TNSM} but to cope with the non stationary behavior of NSPR arrivals, we introduce an additional set of states to describe network load. 

Other methods could be considered to deal with the cyclic nature of NSPR arrivals considered in this paper (e.g., LSTM techniques able to infer the periodic characteristics of the NSPRs arrival process). However, our goal in this paper is to set up a method of dealing with non stationary NSPR arrivals, independently of underlying periodic structures. We test the proposed approach for periodic NSPR arrivals to validate the approach, keeping in mind that the method has to be robust against non stationary traffic variations. This point will be addressed in further studies.

\subsection{Learning framework}
\label{sec:drl_policy}

\subsubsection{Policy} We reuse the  framework introduced in \cite{HA_DRL_TNSM}. We denote by  $\mathcal{A}$ the set of possible actions (namely placing VNFs on nodes)  and by $\mathcal{S}$ the set of all states.  We adopt a sequential placement strategy so that we choose a node $n \in N$ where to place a specific VNF $v \in \{1,...,|V|\}$. The VNFs are sequentially placed  so that placement starts with the VNF $v=1$ and ends for the  VNF $v = |V|$.

 At each time step $t$, given a state $\sigma_t$, the learning agent selects an action $a$ with probability given by  the Softmax distribution given by
\begin{equation}
    \pi_{\theta}(a_{t} = a|\sigma_t) = \frac{e^{Z_{\theta}(\sigma_t,a),}}{\sum_{b \in N}e^{Z_{\theta}(\sigma_t,b)}},
    \label{eq::policy}
\end{equation}
where the  function $Z_{\theta}: \sset \times \aset \rightarrow \mathbb{R}$ yields a real value for each state and action calculated by a Deep Neural Network (DNN) as detailed in Section~\ref{sec::drl_learning}. The notation $\pi_{\theta}$ is used to indicate that policy depends on  $Z_{\theta}$. The control parameter $\theta$ represents the weights in the DNN.

\subsubsection{State representation}

As in \cite{HA_DRL_TNSM}, the \textbf{PSN state} is characterized by the occupancy of servers:   $cap^{cpu} = \{cap^{cpu}_{n}: n \in N\}$, $cap^{ram} = \{cap^{ram}_{n}: n \in N\}$ and $cap^{bw} = \{cap^{bw}_{n} = \sum_{(n,b) \in L}cap^{bw}_{(n,b)}: n \in N\}$. In addition, we keep track of the placement of the outstanding NSPR (under placement) via the vector $\chi = \{\chi_{n} \in \{0,..,|V|\} : n \in N \}$, where $\chi_{n}$ is  the number of VNFs of the current NSPR placed on node $n$.

Accordingly, the \textbf{NSPR state} is a view  of the current placement and is composed of four characteristics, three related to resource requirements (see Table \ref{tab::nspr_parameters} for the notation) of  the current VNF $v$ to be placed: $req^{cpu}_{v}$, $req^{ram}_{v}$ and $req^{bw}_{v} =  \sum_{(v,\bar{b}) \in E}req^{bw}_{(v,\bar{b})}$.In addition, let $m_{v} = |V| - v + 1$ be the number of VNFs of the outstanding  NSPR still  to be placed. 

To deal with the non stationary environment, we  introduce the \textbf{Load state} that represents the network load forecast used to learn the network load variations. It is defined by a set $\rho_j$ of 100 features for each resource $j$ calculated using the network load formula given by Eq.~\eqref{eq:global} as follows : $\rho_j=\{\rho_{j}(t) : t_{a}\leq t \leq t_{a} + 100, t \in \mathbb{N }\}$, where $t_a$ is the simulated time instant in which the current NSPR arrives.   

\subsubsection{Reward function} We reuse the reward function introduced in \cite{HA_DRL_TNSM}. We precisely consider
\begin{equation}
     \small
     r_{t+1} = \left\{\begin{array}{lr}
     0, & \text{if $t < T$ and $a_{t}$ is  successful}\\
     \sum^{T}_{i=0} \delta^{a}_{i+1}\delta^{b}_{i+1}\delta^{c}_{i+1}, & \text{if $t = T$ and $a_{t}$ is successful}\\
     \delta^{a}_{t+1}, & \text{otherwise}
    \end{array}\right.
    \label{eq::reward_function}
\end{equation}
where $T$ is the number of iterations of a training episode and where the rewards  $\delta^{a}_{i+1}$, $\delta^{b}_{i+1}$, and $\delta^{c}_{i+1}$ are defined as follows:
\begin{itemize}
    \item An Action $a_t$ may lead to a successful or unsuccessful placement. We then define the Acceptance Reward value due to action $a_t$ as
\begin{equation}
     \delta^{a}_{t+1} = \left\{\begin{array}{lr}
    100, & \text{if $a_{t}$ is successful, }\\
    -100, & \text{otherwise. }
    \end{array}\right.
    \label{eq::acceptance_signal}
\end{equation}
\item  The Resource Consumption Reward value for the placement of VNF $v$ via action $a_t$ is defined by
\begin{equation}
     \delta^{c}_{t+1}= \left\{\begin{array}{lr}
    \frac{req^{bw}_{(v-1,v)}}{req^{bw}_{(v-1,v)}|P|} = \frac{1}{|P|}, & \text{if $|P|>0$, }\\
    1, & \text{otherwise. }
    \end{array}\right.
    \label{eq::resource_consumption_signal}
\end{equation}
where $P$ is the path used to place VL $(v-1,v)$. Note that a maximum  $\delta^{c}_{t+1} = 1$ is given when $|P|=0$, that is, when VNFs $v-1$ and $v$ are placed on the same server.
\item  The Load Balancing Reward value for the placement of VNF $v$ via $a_t$
\begin{equation}
    \delta^{b}_{t+1} = \frac{cap^{cpu}_{a_t}}{M^{cpu}_{a_{t}}} + \frac{cap^{ram}_{a_t}}{M^{ram}_{a_{t}}}.
    \label{eq::load_balancing_signal}
\end{equation}
\end{itemize}

\subsection{Adaptation of DRL and Introduction of a Heuristic Function
\label{sec:aidedDRL}}

\subsubsection{Proposed Deep Reinforcement Learning Algorithm \label{sec::drl_learning}}

As in \cite{HA_DRL_TNSM}, we use a single thread version of the A3C Algorithm introduced in \cite{a3c}. 
This algorithm relies on two DNNs that are trained in parallel: i) the Actor Network with the parameter $\theta$, which is used to generate the policy $\pi_{\theta}$ at each time step, and ii) the Critic Network with the parameter $\theta_{v}$ which generates an estimate $\nu^{\pi_{\theta}}_{\theta_{v}}(\sigma_t)$ for the  State-value function defined by $$\nu_{\pi}(t|\sigma)=\mathbb{E}_{\pi}\left[\sum^{T-t-1}_{k=0}\gamma^{k} r_{t+k+1} | \sigma_t = \sigma \right],$$
for some discount parameter $\gamma$. 

\begin{figure}[hbtp] 
\centering
\includegraphics[width=\linewidth]{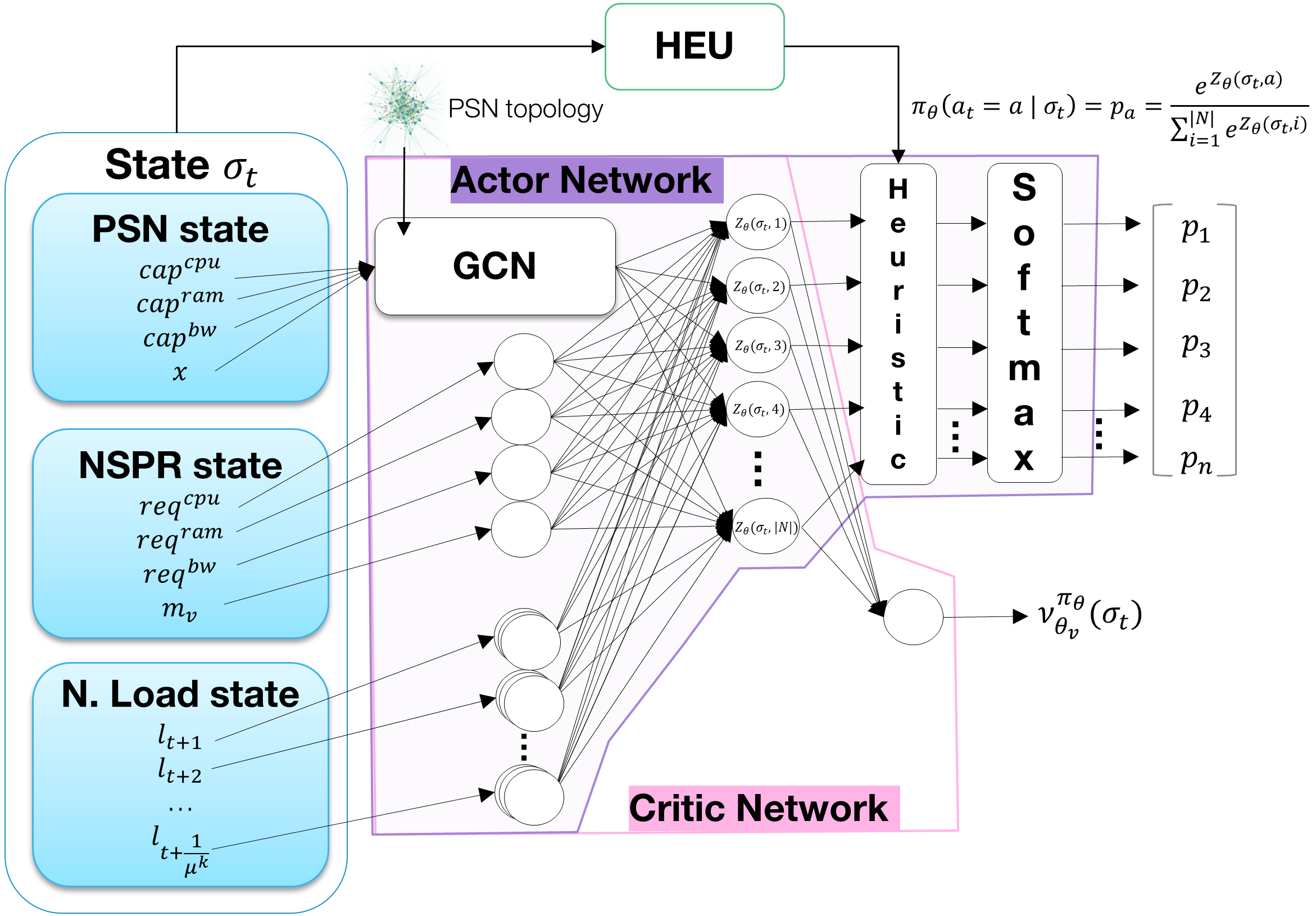}
\caption{Reference framework for the proposed learning algorithms. \label{fig::advantage_actor_critic_architecture_2}}
\end{figure}

As depicted in Fig.~\ref{fig::advantage_actor_critic_architecture_2} both Actor and Critic Networks have almost identical structure. As in \cite{p1}, we use the GCN formulation proposed by \cite{kipf_gcn} to automatically extract advanced characteristics of the PSN. The characteristics produced by the GCN represent semantics of the PSN topology by encoding and accumulating characteristics of neighbour nodes in the PSN graph. The size of the neighbourhood is defined by the order-index parameter $K$. As in \cite{p1}, we consider in the following $K=3$ and perform automatic extraction of 60 characteristics per PSN node. Both the NSPR state and Network Load characteristics are separately transmitted to fully connected layers with 4 and 100 units, respectively. 

The characteristics extracted by both layers and the GCN layer are combined into a single column vector of size $60|N| + 104$ and passed through a full connection layer with $|N|$ units.

In the Critic Network, the outputs are forwarded to a single neuron, which is used to calculate the state-value function estimation $\nu^{\pi_{\theta}}_{\theta_{v}}(\sigma_t)$. In the Actor Network, the outputs represent the values of the function $Z_{\theta}$ introduced in Section \ref{sec:drl_policy}. These values are injected into a Softmax layer that transforms them into a Softmax distribution that corresponds to the policy $\pi_{\theta}$.

During the training phase, at each time step $t$, the  A3C algorithm uses the Actor Network to calculate the policy $\pi_{\theta}(.|\sigma_t)$. An action $a_t$ is sampled using the policy and performed on the environment. The Critic Network is used to calculate the state-value function approximation $\nu^{\pi_{\theta}}_{\theta_{v}}(\sigma_t)$. The learning agent receives then the reward $r_{t+1}$ and next state $\sigma_{t+1}$ from the environment and the placement process continues until a terminal state is reached, that is, until the Actor Network returns an unsuccessful action or until the current NSPR is completely placed. At the end of the training episode, the A3C algorithm updates parameters $\theta$ and $\theta_{v}$ by using the same rules as in \cite{HA_DRL_TNSM}.

\subsubsection{Introduction of a Heuristic Function}
\label{heuristicfunc}
To guide  the learning process, we use as in \cite{HA_DRL_TNSM} the placement heuristic introduced in  \cite{cnsm_2020}. This yields the  HA-DRL algorithm. More precisely, from the reference framework shown in Fig.~\ref{fig::advantage_actor_critic_architecture_2},  we proposed to include in the  Actor Network  the Heuristic layer that calculates a Heuristic Function $H: \sset \times \aset \rightarrow \mathbb{R}$ based on external information provided by the heuristic method,  referred to as HEU.  

Let $Z_{\theta}$ be the function computed by the fully connected layer of the Actor Network that maps each state and action to a real value which is after converted by the Softmax layer into the selection probability of the respective action (see Section \ref{sec:drl_policy}). Let $\bar{a}_{t} = \text{argmax}_{a \in \aset}\,Z_{\theta}(\sigma_t,a)$ be the action with the highest $Z_{\theta}$ value for state $\sigma_{t}$. Let $a^{*}_{t}=HEU(\sigma_t)$ be the action derived by the HEU method at time step $t$ and the preferred action to be chosen. $H(\sigma_t,a^{*}_t)$ is shaped to allow the value of $Z_{\theta}(\sigma_t,a^{*}_t)$ to become closer to the value of $Z_{\theta}(\sigma_t,\bar{a}_t)$.

The aim is to turn $a^{*}_t$ into one of the likeliest actions to be chosen by the policy.

The Heuristic Function is then formulated as
\begin{multline}
     \scriptsize
     H(\sigma_t,a_t) =   \left\{\begin{array}{lr}
     Z_{\theta}(\sigma_t,\bar{a}_{t}) -  Z_{\theta}(\sigma_t,a_t) + \eta, & \text{if $a_{t}=a^{*}_{t}$}\\
     0, & \text{otherwise}
    \end{array}\right.
    \label{eq::heuristic_function}
\end{multline}
where $\eta$ parameter is a small real number. During the training process the Heuristic layer calculates $H(\sigma_t,.)$ and updates the $Z_{\theta}(\sigma_t,.)$ values by using the following equation:
\begin{equation}
    Z_{\theta}(\sigma_t,.) = Z_{\theta}(\sigma_t,.) + \xi H(\sigma_t,.)^{\beta} \label{eq:z_update} 
\end{equation}
The Softmax layer then computes the policy using the modified $Z_{\theta}$. Note the action returned by $a^{*}_{t}$ will have a higher probability to be chosen. The $\xi$ and $\beta$ are parameters used to control how much HEU influence the policy.

\subsection{Implementation Remarks}

\subsubsection{Algorithms considered}
We consider  four learning algorithms based on the reference framework presented in Fig.~\ref{fig::advantage_actor_critic_architecture_2}. 
\begin{itemize}
    \item \textbf{DRL:} It is the pure-DRL algorithm we initially proposed in \cite{HA_DRL_TNSM}. The state representation does not include the network load state and the Actor Network does not contain the Heuristic layer.
    \item \textbf{eDRL:} This algorithm is an enhanced version of DRL in which the state representation includes the network load state and the Actor Network does not contain the Heuristic layer; 
    \item \textbf{HA-DRL:} This algorithm embeds the heuristic but the state representation does not include the network load state while the Actor Network contain the Heuristic layer; 
    \item \textbf{HA-eDRL:} The state representation includes the network load state and the Actor Network contains the Heuristic layer. 
\end{itemize}
\subsubsection{Implementation details}
All resource-related characteristics are normalized to be in  $[0,1]$. This is done by dividing $cap^{j}$ and $req^{j}$, $j \in \{$cpu, ram,bw$\}$, by $\max_{n \in N}M^{j}_{n}$. With regard to the DNNs, we have implemented the Actor and Critic as two independent Neural Networks. Each neuron has a bias assigned. We have used the hyperbolic tangent (tanh) activation for non-output layers of the Actor Network and Rectified Linear Unit (ReLU) activation for all layers of the Critic Network. We have normalized positive global rewards to be in  $[0,10]$. During the training phase, we have considered the policy as a Categorical distribution and used it to sample the actions randomly.

\section{Implementation and  Evaluation Results \label{sec:evaluation}}

In this section, we present the implementation and experiments we conducted to evaluate the proposed algorithms.

\subsection{Implementation Details \& Simulator Settings}

\subsubsection{Experimental setting} We developed a simulator in Python containing: i) the elements of the Network Slice Placement Optimization problem (i.e., PSN and NSPR); ii) the DRL, eDRL, HA-DRL and HA-eDRL algorithms. We used the PyTorch framework to implement the DNNs. 
Experiments were run in a 2x6 cores @2.95Ghz 96GB machine.

\begin{table*}[ht]
\centering
\caption{Network Load Calculation for both NSPR Classes}
\label{tab:network_loads}
\begin{tabular}{@{}ccccccc@{}}
\toprule
\textbf{\begin{tabular}[c]{@{}c@{}}NSPR \\ Classes ($k$)\end{tabular}} &
  \textbf{\begin{tabular}[c]{@{}c@{}}\# of VNFs \\ per NSPR ($|V^{k}|$)\end{tabular}} &
  \textbf{\begin{tabular}[c]{@{}c@{}}CPU requested \\ per VNF ($req^{cpu,k}$)\end{tabular}} &
  \textbf{\begin{tabular}[c]{@{}c@{}}NSPR \\ lifetime ($\frac{1}{\mu^{k}}$)\end{tabular}} &
  \textbf{\begin{tabular}[c]{@{}c@{}}NSPR \\ arrival rate ($\lambda^{k}(t)$)\end{tabular}} &
  \textbf{\begin{tabular}[c]{@{}c@{}}Total CPU \\ capacity ($C^{cpu}$)\end{tabular}} &
  \textbf{\begin{tabular}[c]{@{}c@{}}Network \\ Load ($\rho^{k}(t)$)\end{tabular}} \\ \midrule
Volatile &
  5 &
  25 &
  20 &
  $1.5 \sin^{2}(\frac{\pi t}{96})$&
  6300 &
  $1.5 \sin^{2}(\frac{\pi t}{96})\frac{2500}{6300}$\\
Long term &
  10 &
  25 &
  500 &
  0.02 &
  6300 &
  $0.02\frac{125000}{6300}$ \\ \bottomrule
\end{tabular}
\label{tab:network_load_calculation}
\end{table*}

\subsubsection{Physical Substrate Network Settings} \label{sec::substrate_network_settings}
We consider a PSN that could reflect the infrastructure of an operator as discussed in \cite{farah2}. In this network, three types of DCs are introduced as in Section~\ref{sec:network_model}. Each CDC is connected to three EDCs which are 100 km apart. CDCs are interconnected and connected to one CCP that is 300 km away.
We consider 15 EDCs each one with 4 servers, 5 CDCs each with 10 servers and 1 CCP with 16 servers. The CPU and RAM capacities of each server are 50 and 300 units, respectively. A bandwidth capacity of 100 Gbps is given to intra-data center links inside CDCs and CCP---10Gbps being the case for intra-data center links inside EDCs. Transport links connected to EDCs have 10Gpbs of bandwidth capacity. Transport links between CDCs have 100Gpbs of bandwidth capacity as well for the ones between CDCs and the CCP. 

\subsubsection{Network Slice Placement Requests Settings \label{sec::network_slice_placement_requests_settings}}

We consider NSPRs to have the Enhanced Mobile Broadband (eMBB) setting described in \cite{cnsm_2020}. Each NSPR is composed of 5 VNFs. Each VNF requires 25 units of CPU and 150 units of RAM. Each VL requires 2 Gbps of bandwidth.

\subsection{Algorithms \& Experimental Setup }\label{sec:algorithms_tested}

\subsubsection{Training Process \& Hyper-parameters}
We consider a training process with maximum duration of 55 hours for the considered algorithms. We perform seven independent runs of each algorithm to assess their average and maximal performance in terms of  metrics introduced below (see Section \ref{sec:ev_metrics}). After performing Hyper-parameter search, we set the learning rates for the Actor and Critic networks of DRL and HA-DRL algorithms to $\alpha = 5 \times 10^{-5}$ and $\alpha' = 1.25 \times 10^{-3}$, respectively. For eDRL and HA-eDRL, we set the learning rates for the Actor and Critic networks to $\alpha = 5.7 \times 10^{-5}$ and $\alpha' = 1.4 \times 10^{-3}$, respectively. We implement  four versions of HA-DRL and HA-eDRL agents, each with a different value for the $\beta$ parameter of the heuristic function formulation (see Section \ref{heuristicfunc}). We set in addition the parameters $\xi = 1$ and $\eta = 0$.

\subsubsection{Network load calculation}\label{sec:network_loads}

We implement the network load model introduced in Section \ref{sec:network_loads} considering two NSPR classes: i) a volatile class (referred to as Volatile), with a dynamic arrival rate and ii) a static class (referred to as Long term), with a static arrival rate. Table~\ref{tab:network_loads} presents the  network load models proposed for both classes. We consider that each simulation time unit corresponds to 15  minutes in reality. We set the period for the network load function to 96 simulation time units, i.e., one  day. The network global network load  $\rho(t)$ varies then between 0.3 and 1.0.  

\subsection{Evaluation Metrics \label{sec:ev_metrics}}
To characterize the performance of the placement algorithms, we consider two performance metrics: 
\begin{enumerate}
\item \textbf{Global Acceptance Ratio (GAR):} The Acceptance Ratio of the different tested algorithms during training computed after each arrival as follows: $\frac{\mathrm{\# accepted \; NSPRs}}{\mathrm{\# arrived \; NSPRs}}$. This metric is used to evaluate the accumulated acceptance ratio of the different algorithms as the learning process progresses.
\item \textbf{Acceptance Ratio per training phase (TAR):} The Acceptance Ratio obtained in each training phase, i.e., each part of the training process, corresponding to $10^4$ NSPR arrivals or $10^4$ episodes. It is calculated as follows: $\frac{\mathrm{\# accepted \; NSPRs}}{\mathrm{10^4}}$. This metric allows us to better observe the evolution of algorithm performance over time since it measures algorithm performance in independent parts (phases) of the training process without accumulating the performance of previous training phases.
\end{enumerate}

\begin{figure}[hbtp] 
\centering
\begin{subfloat}[Maximal performance in 3 simulated days.]
{ \scalebox{.4}{ {\includegraphics{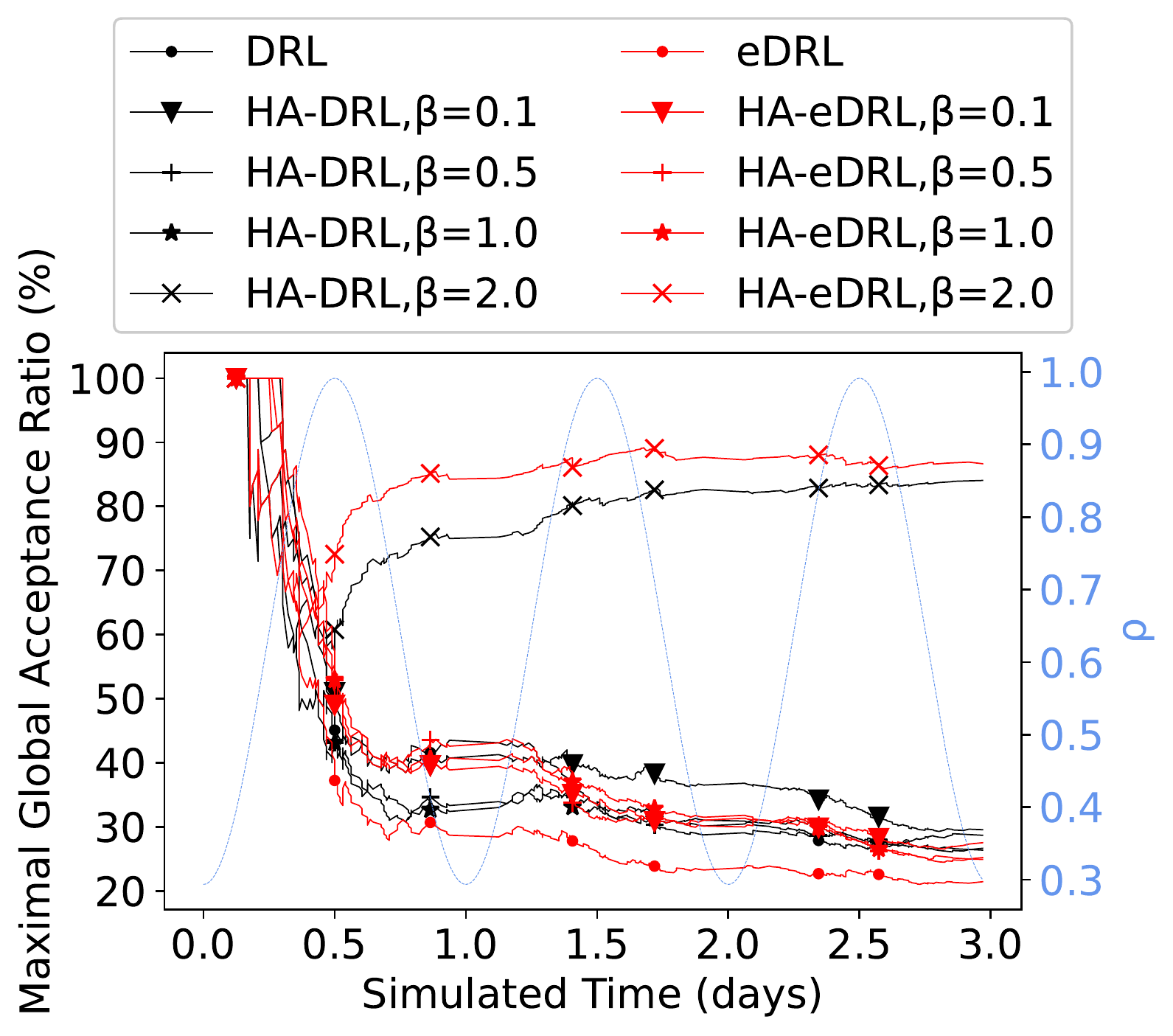}\label{fig:acc_ratio_init_max}}}}
\end{subfloat}

\begin{subfloat}[Maximal performance in 4500 simulated days.]
{ \scalebox{.4}{{\includegraphics{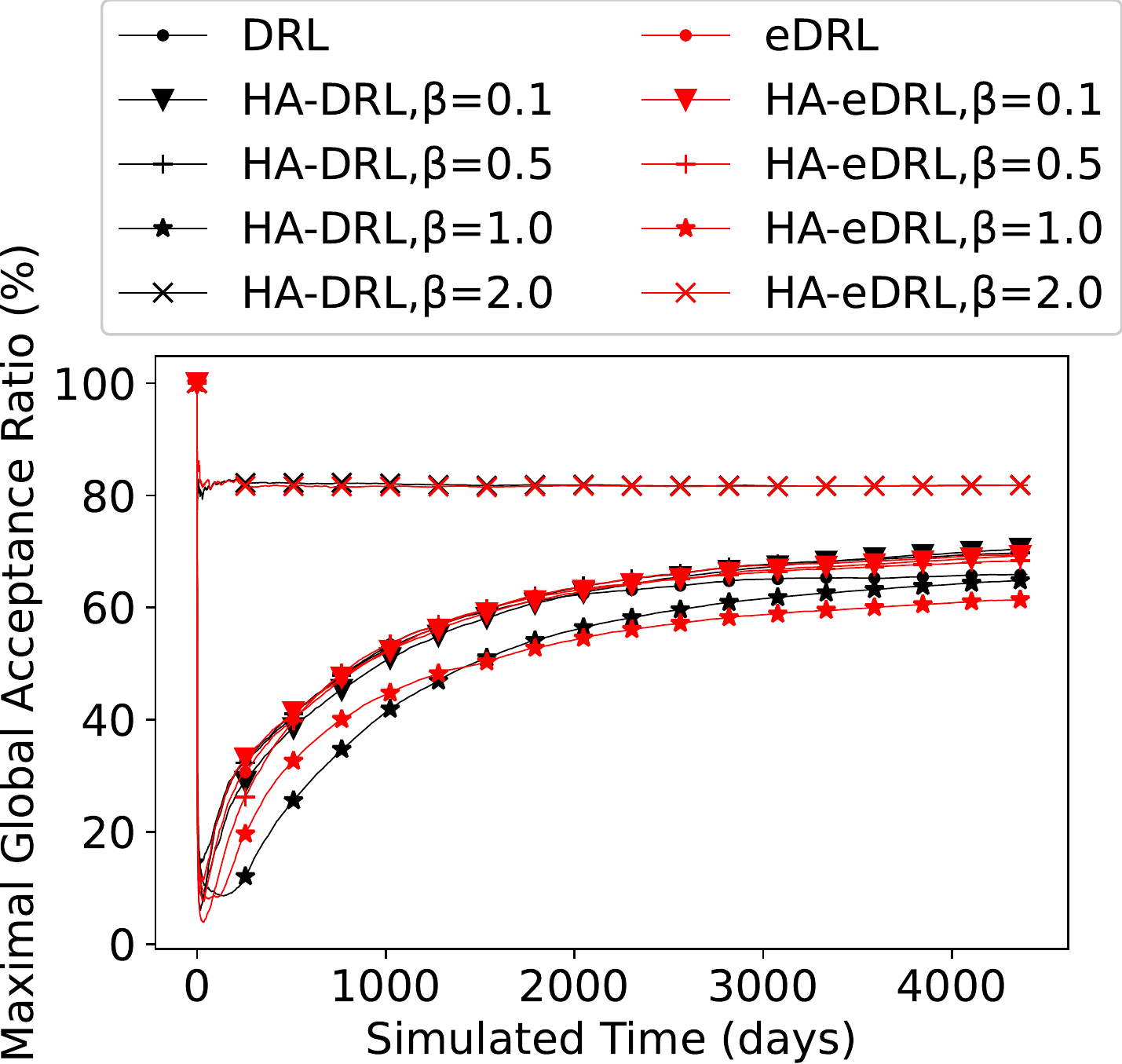}\label{fig:total_max}}}}
\end{subfloat}
\caption{Maximal Global Acceptance ratio results.}
\label{fig:ars_global_1}
\end{figure}

\begin{figure}[hbtp]
\begin{subfloat}[Average performance in 3 simulated days. ]
{ \scalebox{.4}{{\includegraphics{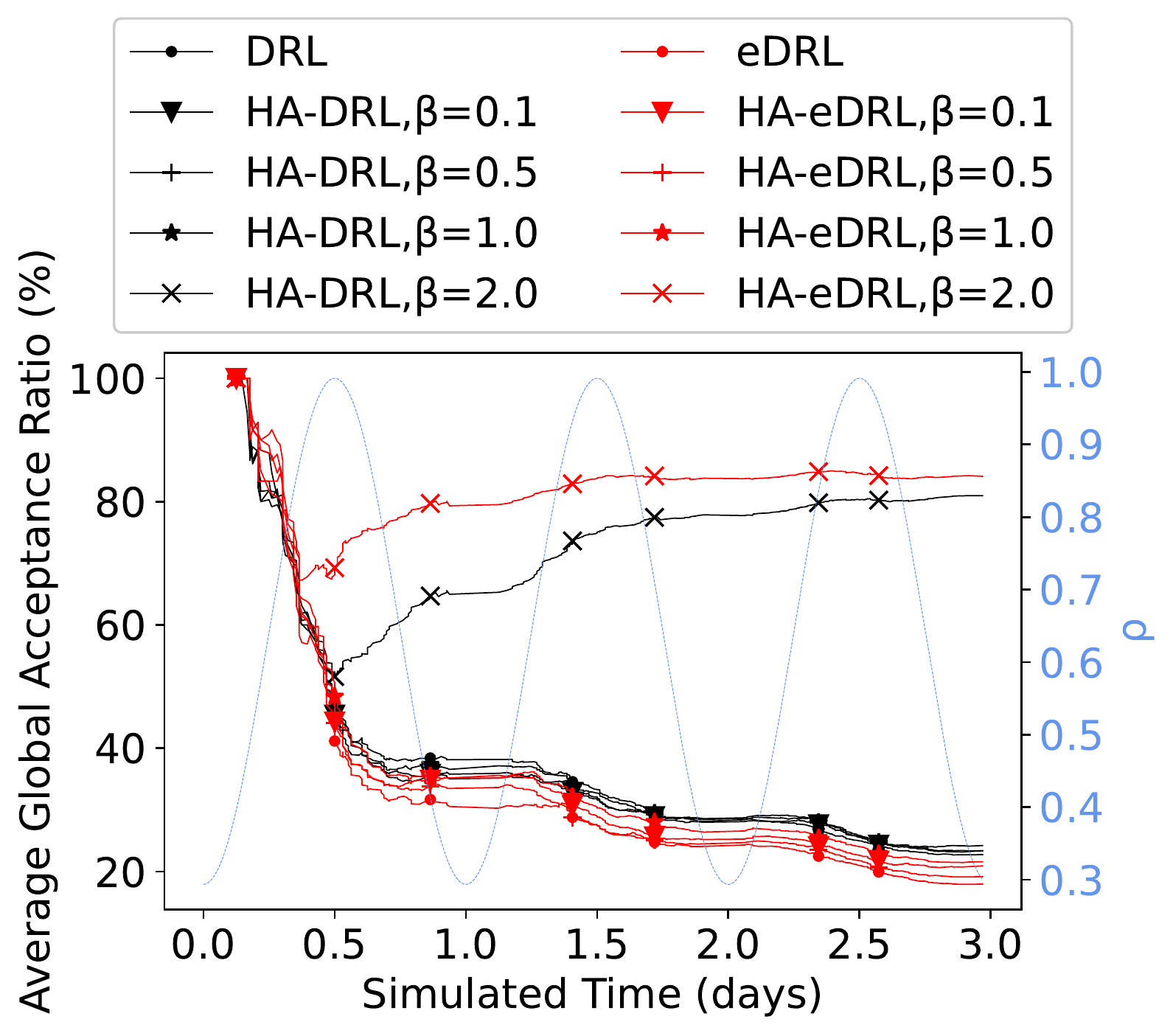}}}\label{fig:init_avg}}
\end{subfloat}

\begin{subfloat}[Average performance in 4500 simulated days.]
{ \scalebox{.4}{\includegraphics{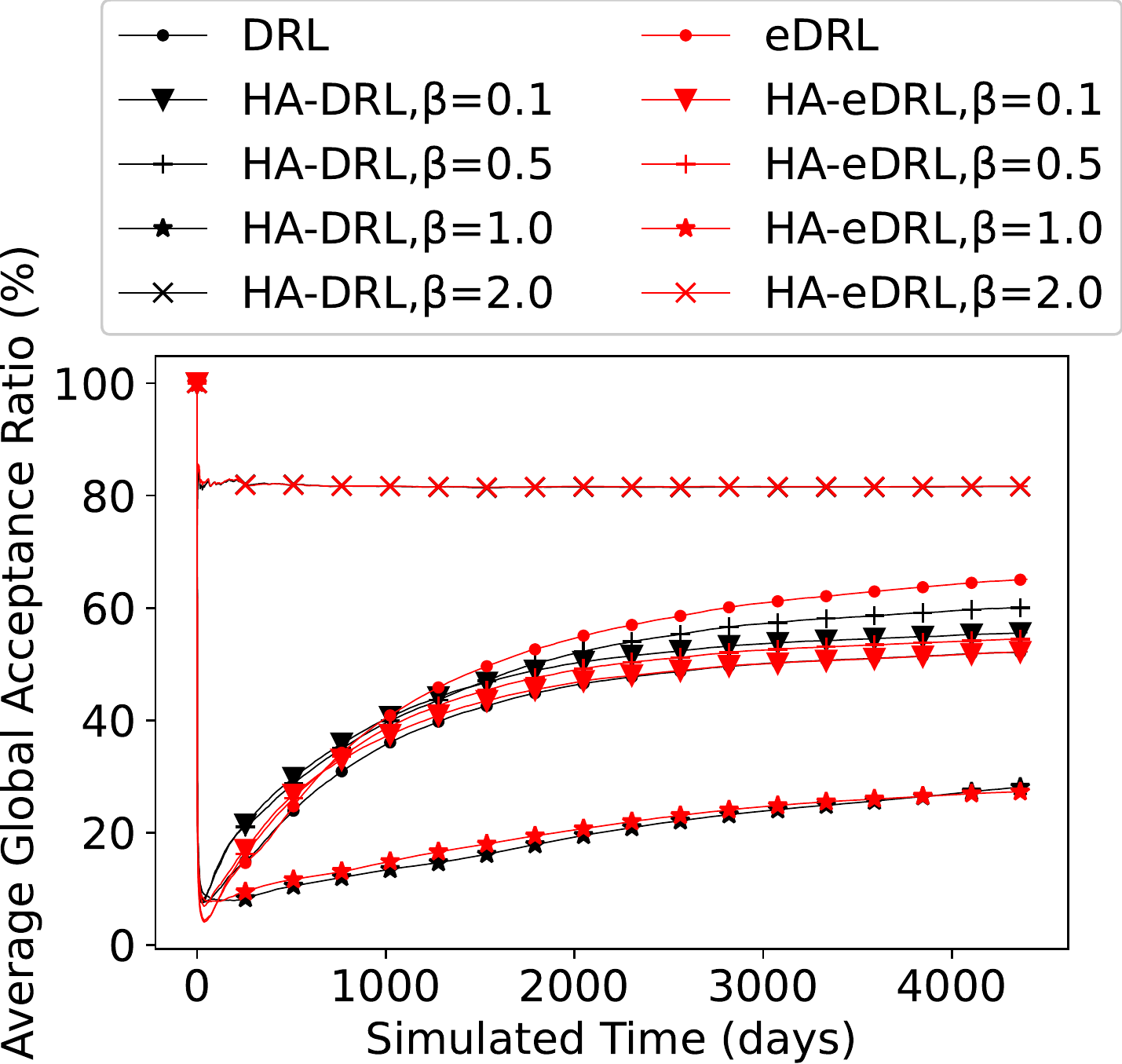}}    \label{fig:total_avg}}
\end{subfloat}
\caption{Average Global Acceptance ratio results.}
\label{fig:ars_global_2}
\end{figure}

\subsection{Global Acceptance Ratio Evaluation \label{sec:global_ev}}

Fig.~\ref{fig:ars_global_1} and \ref{fig:ars_global_2} show the progression of the  Global Acceptance Ratio (GAR) over time  for the considered algorithms. Fig.~\ref{fig:acc_ratio_init_max} and Fig.~\ref{fig:init_avg} show that after 3 simulated days, the average and maximal GARs of HA-DRL and HA-eDRL with $\beta = 2.0$ are convergent and higher than 80\% while for the other algorithms, the GARs remain between 20\% and 40\% and are not  stable. Fig.~\ref{fig:total_max} and Fig~\ref{fig:total_avg} exhibit the progression of the GARs over a longer simulated time horizon of 4500 simulated days. 

We observe that while GARs of HA-DRL and HA-eDRL with $\beta = 2.0$ remain stable after the convergence reached in the first 3 simulated days, for the other algorithms, these performance metrics start to stabilize only after 2000 simulated days. Table~\ref{tab:eval_results} shows that both the maximal and average final GARs, i.e., maximal and average GARs achieved at the end of training, are close to 82\% for both HA-DRL and HA-eDRL with $\beta = 2.0$. For the other algorithms, maximal final GARs are between 61\% and 71\% and average final GARs are between 52\% and 65\%, except for HA-DRL and HA-eDRL with $\beta=1.0$. Table~\ref{tab:eval_results} also shows that maximal and average final GARs of HA-DRL algorithms are generally higher than the equivalent HA-eDRL versions, the gap being never higher than 6\%.

\begin{table}[hbtpt]
\centering
\caption{Summary of evaluation results}
\label{tab:eval_results}
\begin{tabular}{ccccc}
\hline
Algorithm &
  \begin{tabular}[c]{@{}c@{}}Maximal\\ Final\\ GAR (\%)\end{tabular} &
  \begin{tabular}[c]{@{}c@{}}Average \\ Final \\ GAR. (\%)\end{tabular} &
  \begin{tabular}[c]{@{}c@{}}Maximal\\ Final\\ TAR (\%)\end{tabular} &
  \begin{tabular}[c]{@{}c@{}}Average\\ Final\\ TAR (\%)\end{tabular} \\ \hline
DRL           & 65.90 & 52.15 & 77.69 & 55.82 \\
HA-DRL,$\beta=0.1$  & 70.56 & 55.57 & 81.45 & 59.83 \\
HA-DRL,$\beta=0.5$  & 69.80 & 60.12 & 73.56 & 65.33 \\
HA-DRL,$\beta=1.0$  & 64.89 & 28.23 & 72.14 & 41.69 \\
HA-DRL,$\beta=2.0$  & 81.84 & 81.62 & 83.91 & 82.21 \\
eDRL          & 69.36 & 65.10 & 77.39 & 58.61 \\
HA-eDRL,$\beta=0.1$ & 69.25 & 52.22 & 76.34 & 48.45 \\
HA-eDRL,$\beta=0.5$ & 68.40 & 54.49 & 71.91 & 57.57 \\
HA-eDRL,$\beta=1.0$ & 61.46 & 27.33 & 67.05 & 33.10 \\
HA-eDRL,$\beta=2.0$ & 81.82 & 81.68 & 81.37 & 81.95 \\ \hline
\end{tabular}
\end{table}

The above  results show that HA-DRL and HA-eDRL with $\beta = 2.0$ exhibit more robust performance with faster convergence and handle better network load variations than the other evaluated algorithms, notably when compared with the classical DRL approach.

Indeed, when setting $\beta = 2.0$, the Heuristic Function computed on the basis of the HEU algorithm has strong influence on the actions chosen by the agent. Since the HEU algorithm often indicates a good action, this strong influence of the heuristic function helps the algorithms to become stable more quickly.

The addition of network load related features to the state observed by the eDRL algorithm helps  improve the maximal and average final GAR when compared with DRL. However, this improvement is less significant than the improvement brought by the Heuristic Function acceleration as the eDRL algorithm does not achieve the same fast convergence and robustness than HA-DRL and HA-eDRL with $\beta=2.0$.

\begin{figure*}[ht]
\begin{subfloat}[Training Phases 0--16.]
 {\includegraphics[width=.32\linewidth]{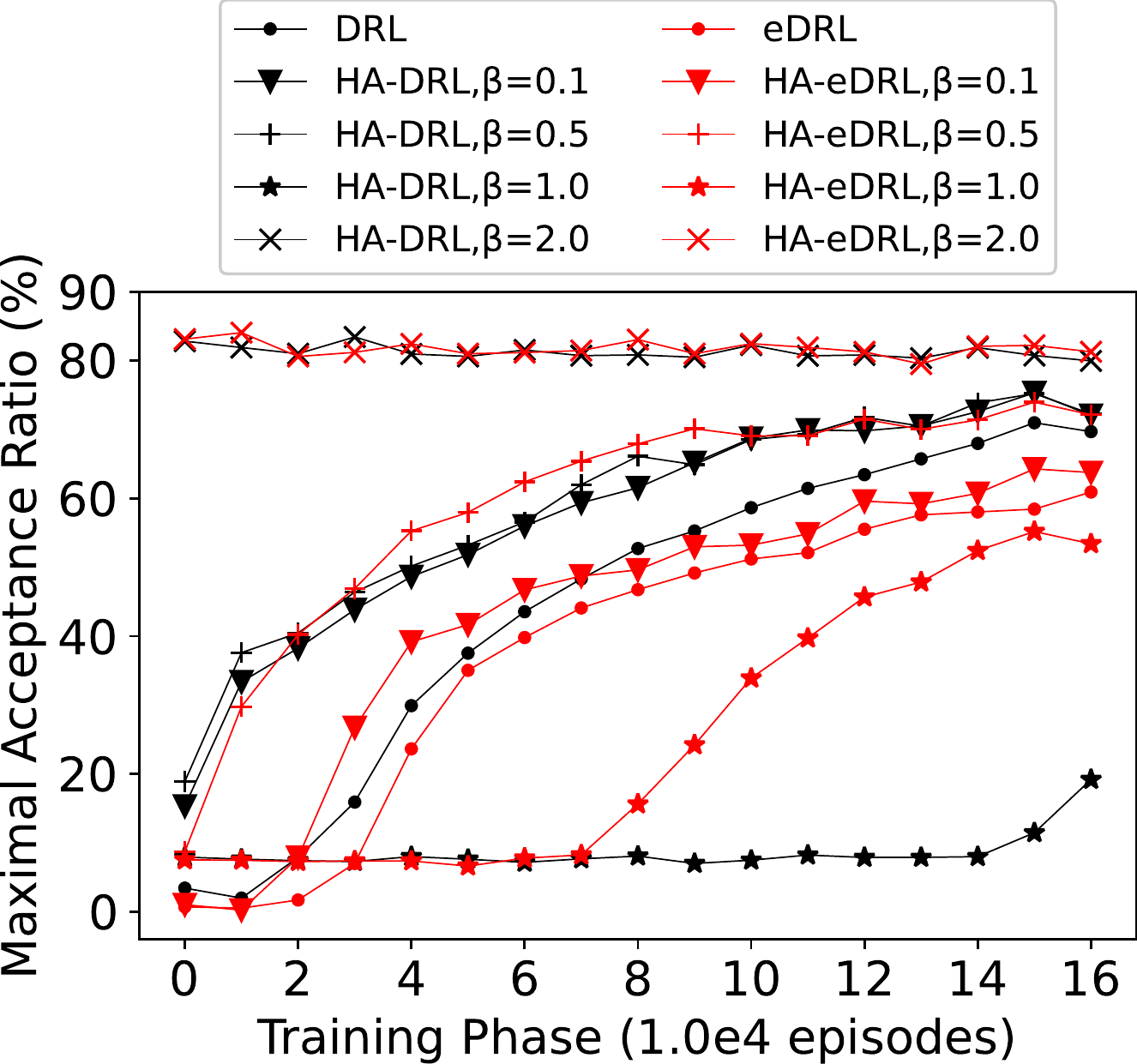}\label{fig:ar_max_1-16}}
\end{subfloat}
\begin{subfloat}[Training Phases 17--33]
 {\includegraphics[width=.32\linewidth]{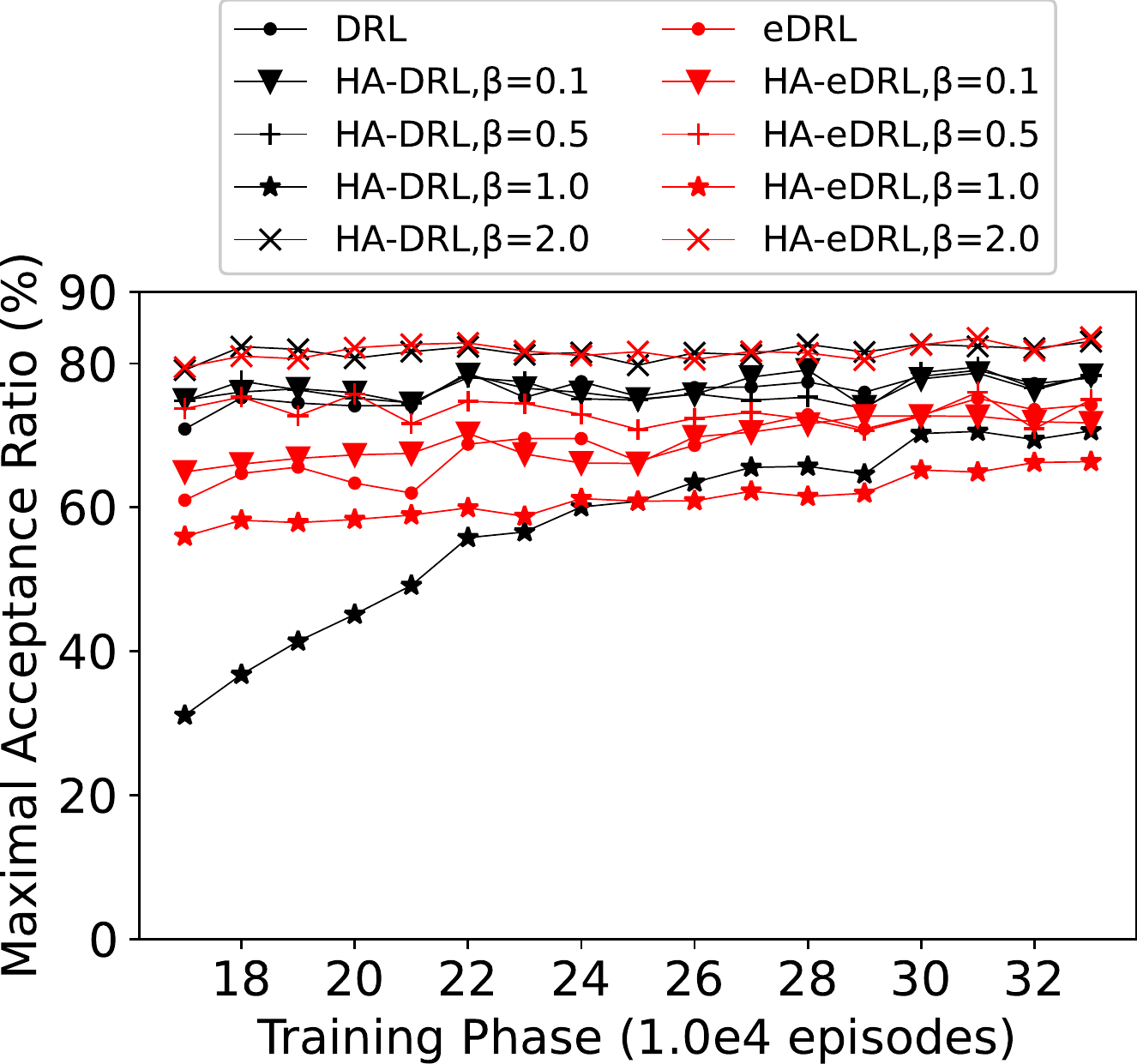}\label{fig:ar_max_17-34}}
\end{subfloat}
\begin{subfloat}[Training Phases 34--50]
 {\includegraphics[width=.32\linewidth]{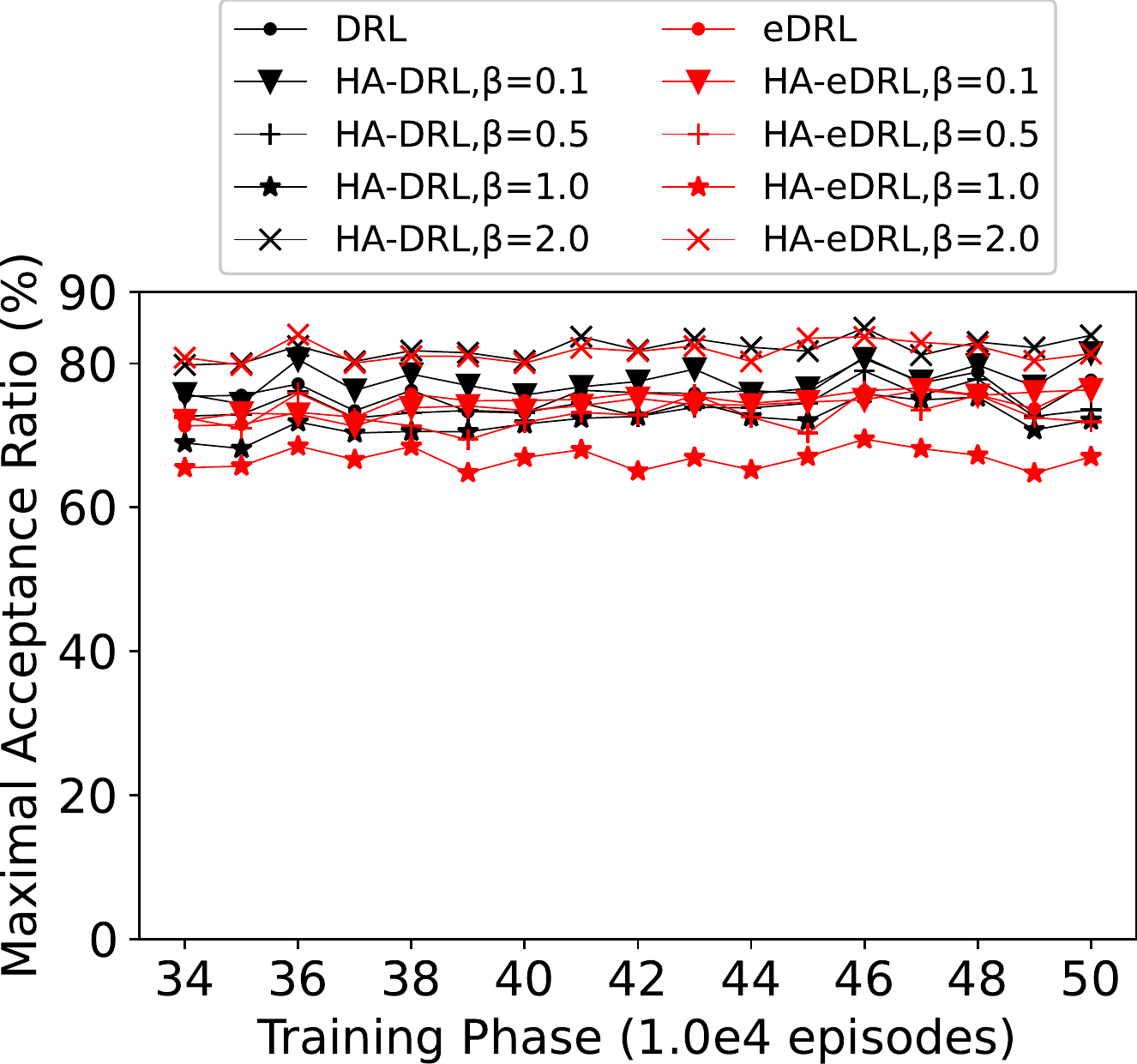}\label{fig:ar_max_34-50}}
\end{subfloat}
\caption{Maximal Acceptance Ratio per training phase results.}
\label{fig:ar_vs_tr_phase_1}
\end{figure*}

\begin{figure*}[ht]
\begin{subfloat}[Training Phases 0--16.]
 {\includegraphics[width=.32\linewidth]{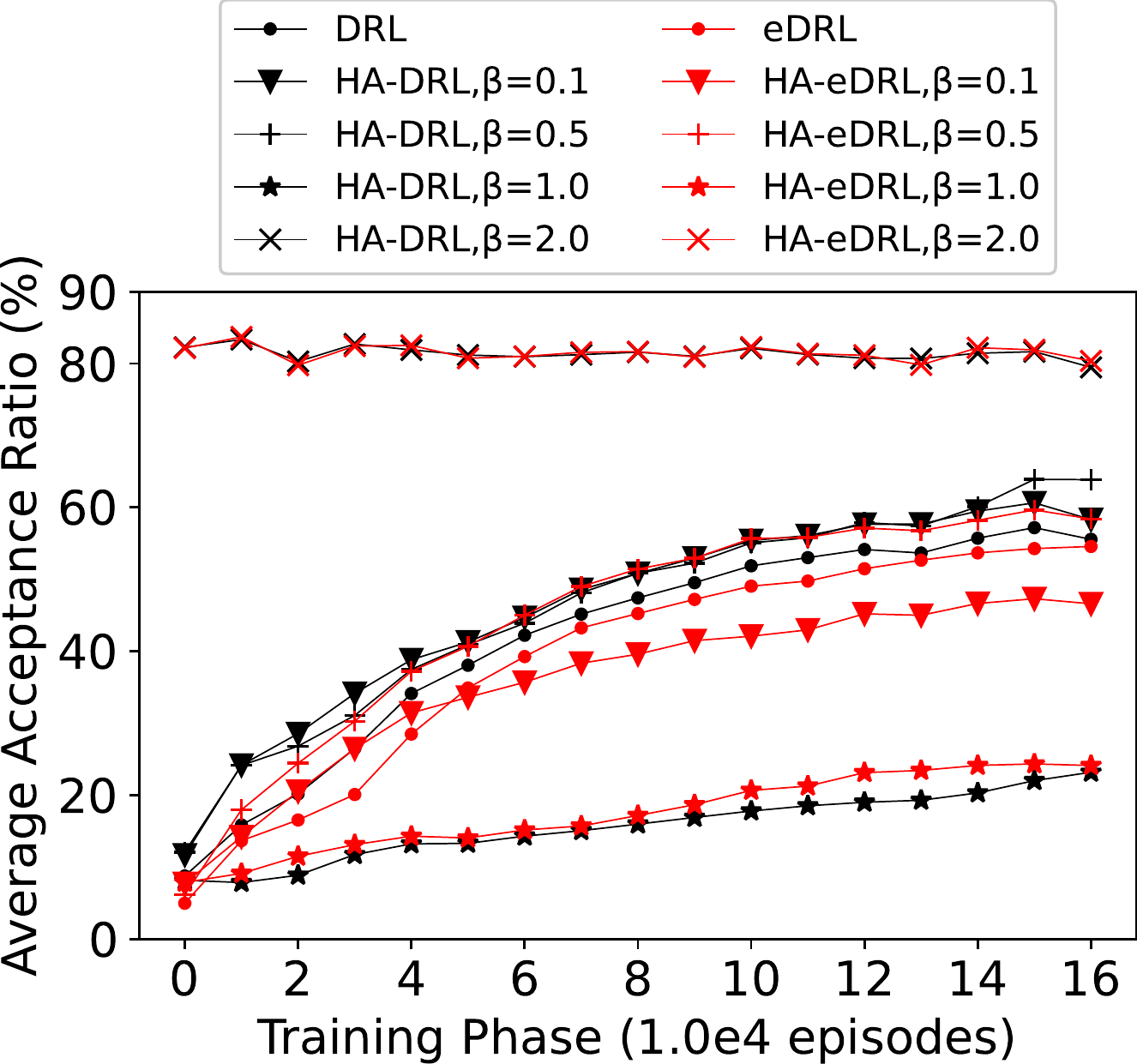}\label{fig:ar_avg_1-16}}
\end{subfloat}
\begin{subfloat}[Training Phases 17--33]
 {\includegraphics[width=.32\linewidth]{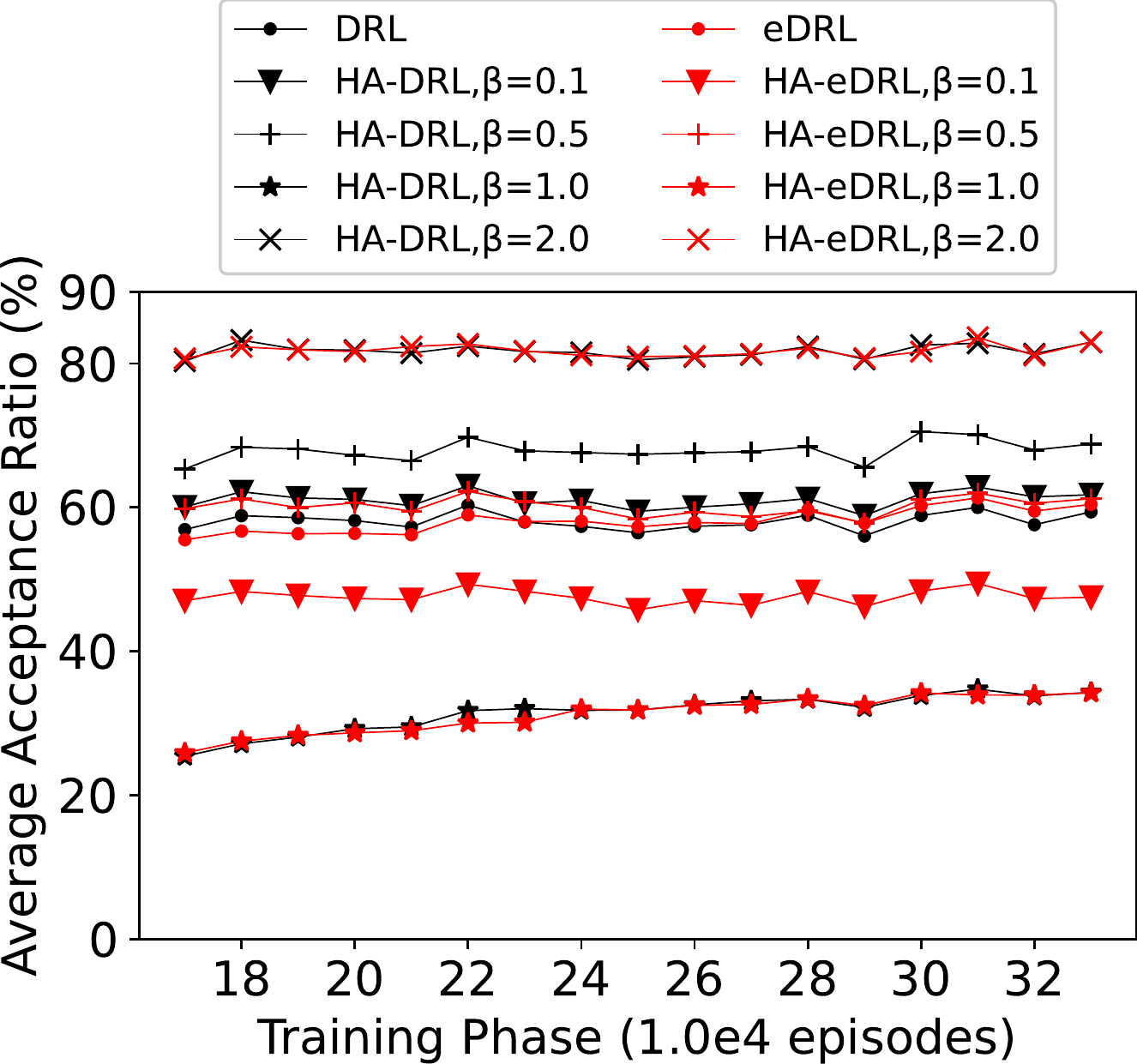}\label{fig:ar_avg_17-34}}
\end{subfloat}
\begin{subfloat}[Training Phases 34--50]
 {\includegraphics[width=.32\linewidth]{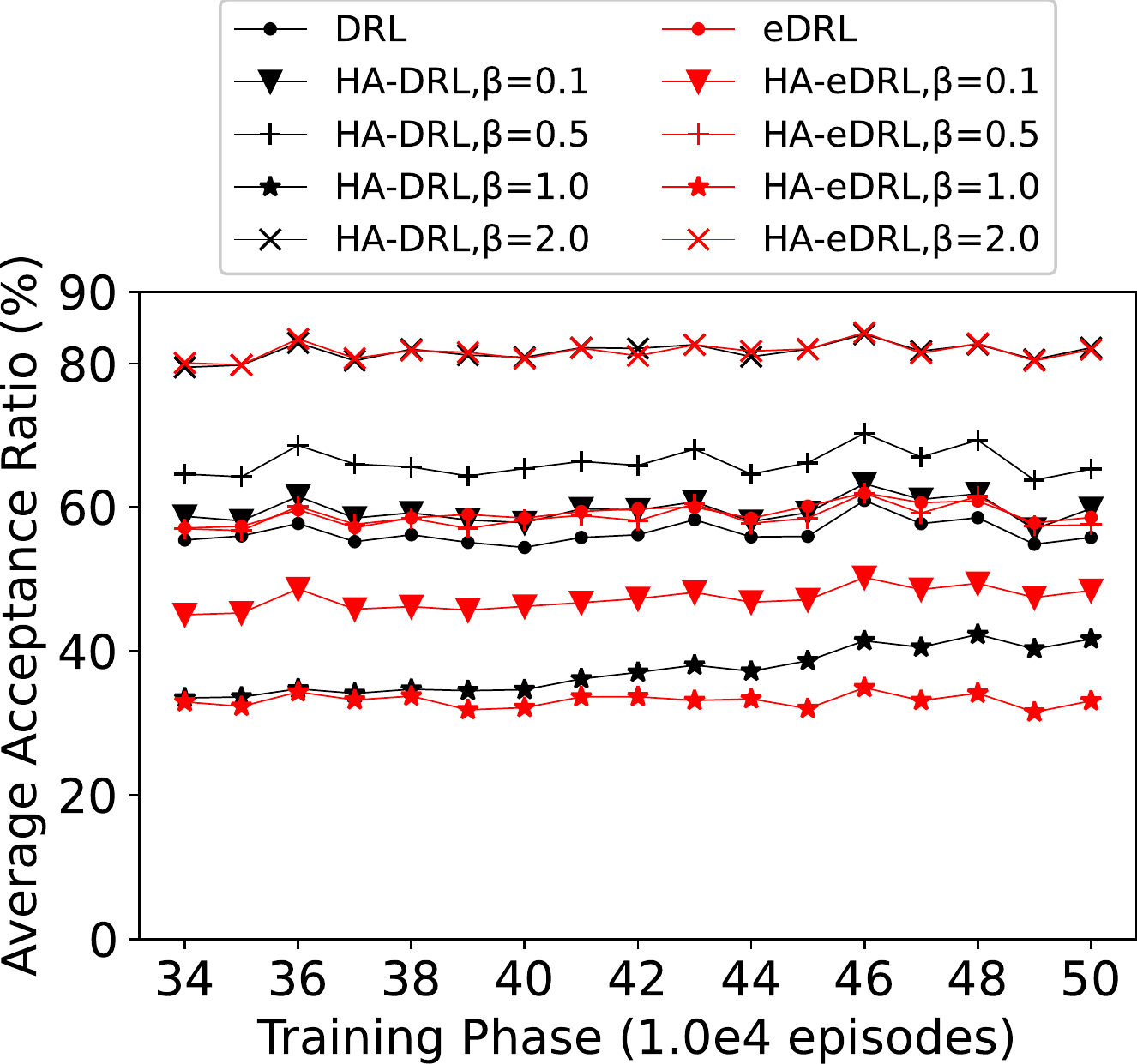}\label{fig:ar_avg_34-50}}
\end{subfloat}
\caption{Average Acceptance Ratio per training phase results.}
\label{fig:ar_vs_tr_phase_2}
\end{figure*}
\begin{figure}[hbtp]
\center
\begin{subfloat}[Maximal performance for volatile NSPRs.]
{\scalebox{.37}{\includegraphics{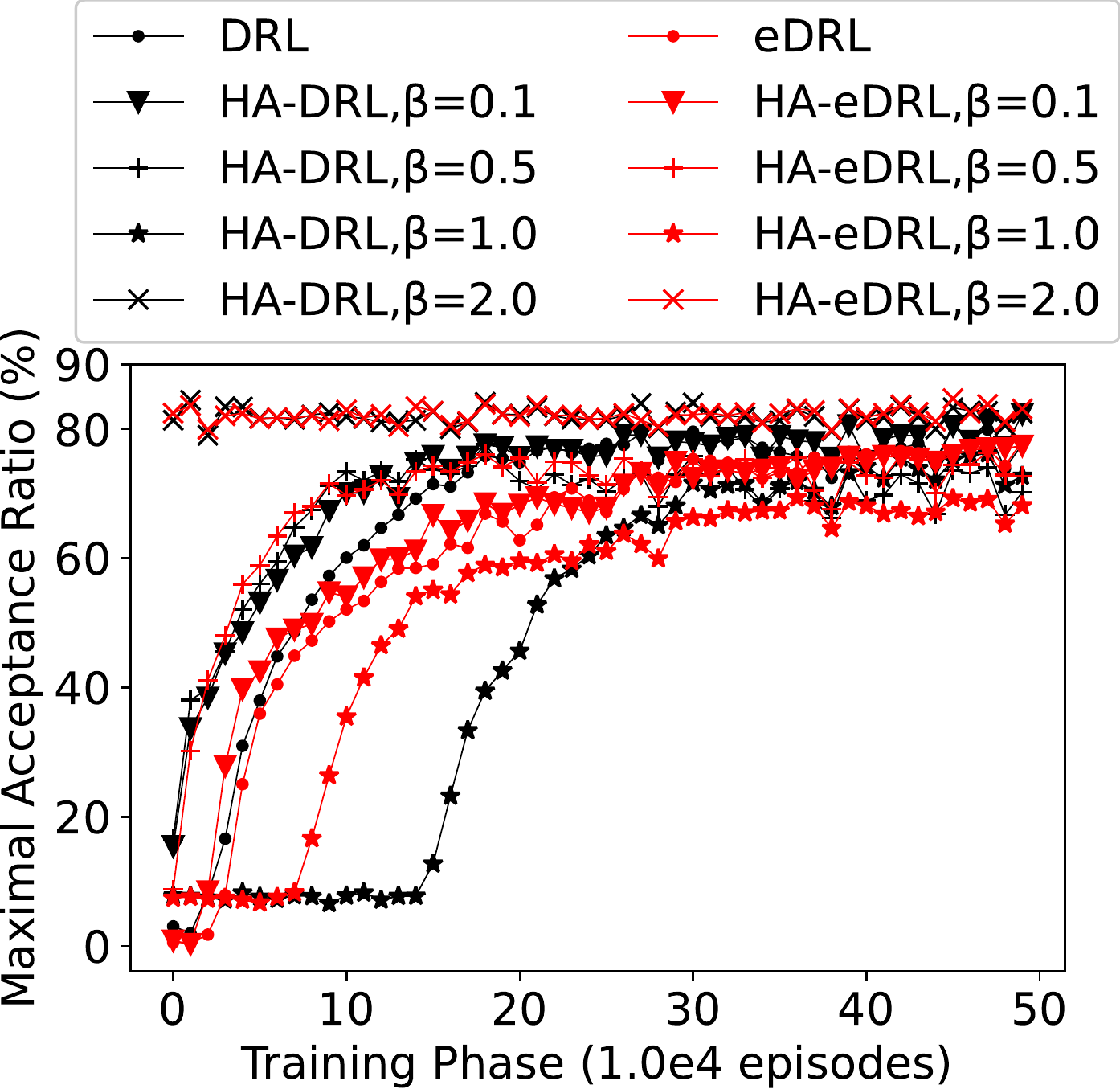}}\label{fig:ar_volatile}}
\end{subfloat}

\begin{subfloat}[Maximal performance for longterm NSPRs.]
{\scalebox{.37}{\includegraphics{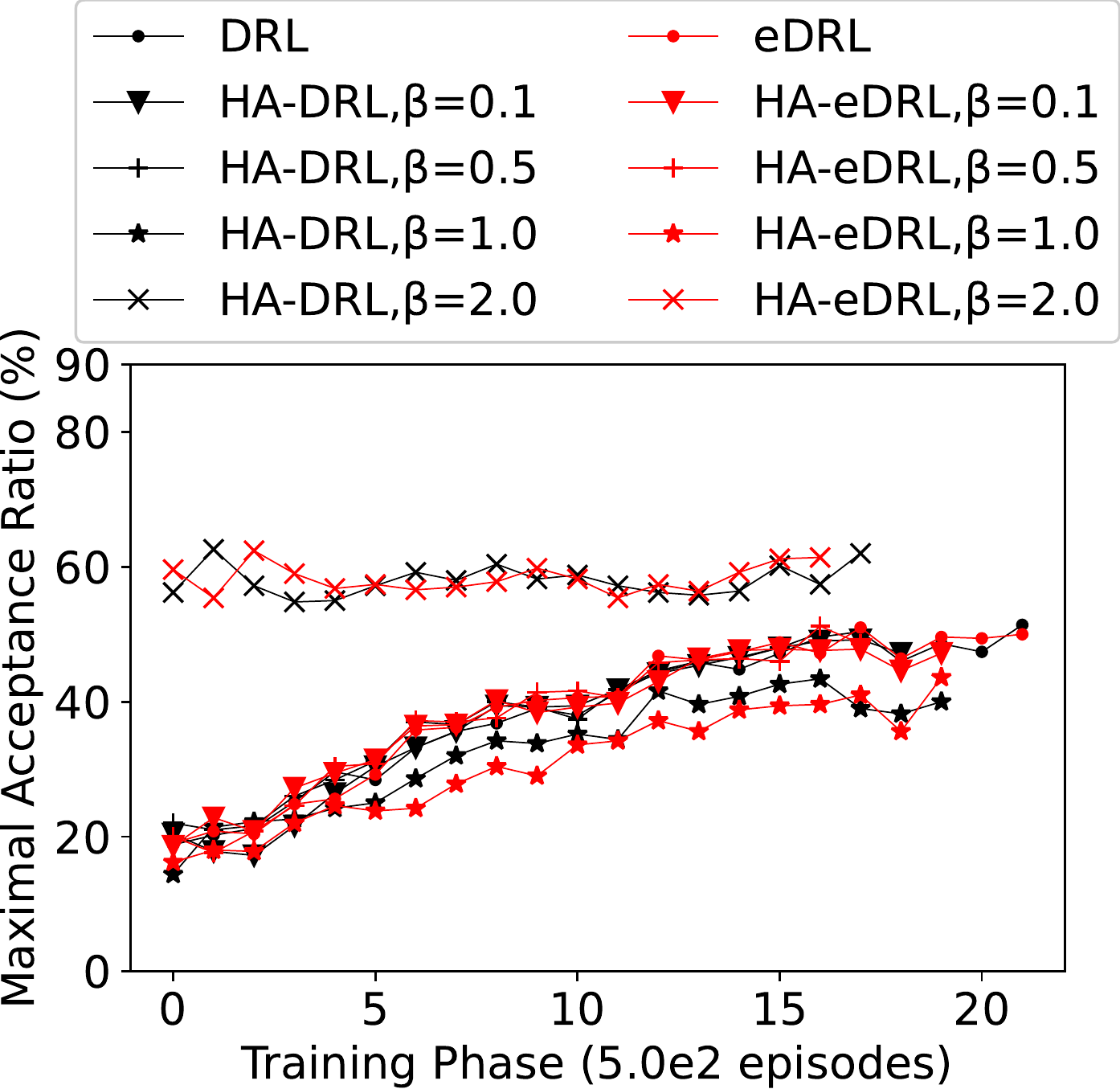}}\label{fig:ar_longterm}}
\end{subfloat}
\caption{Acceptance Ratio per training phase results (per NSPR class).}
\label{fig:ar_vs_tr_phase_3}
\end{figure}

\subsection{Acceptance Ratio per Training Phase Evaluation \label{sec:tar_ev}}

Fig.~\ref{fig:ar_vs_tr_phase_1} and Fig.~\ref{fig:ar_vs_tr_phase_2} show that all algorithms need at least 16 training phases to reach a convergent maximal and average Acceptance Ratio per Training Phase (TAR), except for HA-DRL and HA-eDRL with $\beta=2.0$ which need only one training phase (see Fig~\ref{fig:ar_max_1-16} and Fig~\ref{fig:ar_avg_1-16}). The fast convergence of HA-DRL and HA-eDRL with $\beta=2.0$ is due to the strong influence of the Heuristic Function in the choice of actions as explained in Section \ref{sec:global_ev}. Fig.~\ref{fig:ar_max_1-16} and Fig.~\ref{fig:ar_max_17-34} illustrate that algorithms HA-DRL and HA-eDRL with $\beta=0.5$, HA-DRL with $\beta=0.1$ and DRL need around 16 phases to reach a convergent maximal TAR while Fig.~\ref{fig:ar_max_17-34} and Fig.~\ref{fig:ar_max_34-50} demonstrate that eDRL, HA-eDRL with $\beta \in \{0.1, 1.0\}$ and HA-DRL with $\beta=1.0$ need around 34 phases.  Finally, Fig.~\ref{fig:ar_avg_1-16} and Fig.~\ref{fig:ar_avg_17-34} show that HA-DRL and HA-eDRL with $\beta \in \{0.1, 0.5\}$, eDRL, and DRL need around 16 phases to obtain a convergent average TAR while Fig.~\ref{fig:ar_avg_17-34} and Fig.~\ref{fig:ar_avg_34-50} show that HA-DRL and HA-eDRL with $\beta = 1.0$ need around 34 phases.
 
Table \ref{tab:eval_results} shows that the only algorithms that reach a maximal and average final TAR (i.e, maximal and average TAR achieved at the end of training) higher than 80\% are HA-DRL and HA-eDRL with $\beta=2.0$. HA-DRL with $\beta=0.1$ also attains a maximal final TAR higher than 80\% but its average final TAR is less than 60\%. HA-DRL with $\beta=0.5$ has average final TAR higher than 65\%, but its maximal final TAR is around 74\%.  In addition, Table \ref{tab:eval_results} shows that: HA-DRL algorithms have maximal final TAR generally higher than the equivalent HA-eDRL versions, the gap being never higher than 6\%; eDRL and DRL have equivalent maximal final TAR but eDRL has average final TAR 3\% higher. Fig.~\ref{fig:ar_longterm} and Fig.~\ref{fig:ar_volatile} show that all algorithms have better maximal and average TAR on volatile NSPRs than on long-term NSPRs. This difference is arguably related to the number of arrivals of each requests and, therefore, the amount of training performed on each class of requests, since many more volatile requests arrive to be placed during the simulated period than long-term requests. These results reinforce the conclusion introduced in Section~\ref{sec:global_ev}. The maximal TAR progression results (i.e., Fig.~\ref{fig:ar_vs_tr_phase_1}) allow us to observe that in the best case, if a large number of training phases is given, all the algorithms attain a good performance. 

However, the average TAR progression results (see Fig.~\ref{fig:ar_vs_tr_phase_2}) show that only HA-DRL and HA-eDRL with $\beta=2.0$ have robust performance. These algorithms are also the only one to have quick convergence being, among those evaluated, the most adapted to be used in practice.

\section{Conclusion \label{sec:conclusion}}
We have introduced a DRL-heuristic algorithm that supports the variations of network load in the placement of network slice requests  as a follow-up of our work in \cite{HA_DRL_TNSM}. We  have considered four families of algorithms, pure-DRL and eDRL techniques and their variants combined with a heuristic whose efficiency was investigated in a previous work \cite{cnsm_2020}. The influence of the heuristic on the DRL algorithm can be tuned by means of a parameter (namely, parameter $\beta$ (introduced in Section~\ref{heuristicfunc}). We assume network load has periodic fluctuations. We have shown how introducing  network load states into the DRL algorithm together with a combination with the heuristic function yields very good results in terms of GAR and TAR which are stable in time. This study aims at proving that coupling DRL and heuristic functions yields good and stable results even under non stationary conditions. 

The next step is to study the behavior of the proposed algorithm in case of an unpredictable network load disruption.

%

\section*{Acknowledgment}

This work has been performed in the framework of 5GPPP MON-B5G project (www.monb5g.eu). The experiments were conducted using Grid'5000, a large scale testbed by Inria and Sorbonne University (www.grid5000.fr).

%

\bibliographystyle{IEEEtran}
\bibliography{IEEEabrv,my_bib}

\flushend

\end{document}